\documentclass[aps,prd,amsmath,amssymb,superscriptaddress,nofootinbib,10pt,showpacs,twocolumn]{revtex4-2}

\usepackage{newtxtext,newtxmath}

\usepackage[T1]{fontenc}

\DeclareRobustCommand{\VAN}[3]{#2}
\let\VANthebibliography\thebibliography
\def\thebibliography{\DeclareRobustCommand{\VAN}[3]{##3}\VANthebibliography}

\usepackage{graphicx}	
\usepackage{amsmath}	
\usepackage{dcolumn}
\usepackage{bm}
\usepackage{hyperref}
\usepackage{xcolor}

\def\bea{\begin{eqnarray}}
\def\eea{\end{eqnarray}}
\def\be{\begin{equation}}
\def\ee{\end{equation}}

\def\xspace{\,}

\def\LCDM{$\Lambda$CDM\xspace }

\def\rex{\rho_{ex} }

\def\nbodykit{\textsc{N-BodyKit}\xspace}

\def\apj{ApJ}
\def\apjl{ApJL}

\def\mnras{MNRAS}

\newcommand{\Mpc}{\,h^{-1}\mathrm{Mpc}}
\newcommand{\iMpc}{\,h \,\mathrm{Mpc}^{-1}}
\newcommand{\Mass}{\,h^{-1}M_{\odot}}

\newcommand{\fastpm}{\textsc{FastPM}\xspace}

\newcommand{\ngenicaddress}{https://www.h-its.org/2014/11/05/ngenic-code/}

\begin{document}

\preprint{APS/123-QED}


\title[Halo mass function evolution in bump cosmologies]{Impact of a Rapid Diluted Energy Density on the halo mass function} 

\author{Dante V. Gomez-Navarro}\email{ dantegomezn@gmail.com}
\affiliation{Instituto de F\'isica, Universidad Nacional Aut\'onoma de M\'exico, Cd. de M\'exico C.P. 04510, M\'exico.}

\author{Alejandro Aviles}
\email{avilescervantes@gmail.com}
\affiliation{Departamento de F\'isica, Instituto Nacional de Investigaciones Nucleares,
Apartado Postal 18-1027, Col. Escand\'on, Ciudad de M\'exico,11801, M\'exico.}
\affiliation{Consejo Nacional de Ciencia y Tecnolog\'ia, Av. Insurgentes Sur 1582,
Colonia Cr\'edito Constructor, Del. Benito Ju\'arez, 03940, Ciudad de M\'exico, M\'exico.}

\author{Axel de la Macorra}\email{macorra@fisica.unam.mx}
\affiliation{Instituto de F\'isica, Universidad Nacional Aut\'onoma de M\'exico, Cd. de M\'exico C.P. 04510, M\'exico.}

\begin{abstract}
We study dark energy cosmological models, extensions of the standard model of particles, characterised by having an extra relativistic energy density at very early times, and that rapidly dilute after a phase transition occurs. These models generate well localized features (or bumps) in the matter power spectrum for modes crossing the horizon during the phase transition epoch. This is because the presence of the additional energy component enhances the growth of matter fluctuations. Instead of considering a particular model, we focus on a parametric family of Gaussian bumps in the matter power spectrum, which otherwise would be a $\Lambda$CDM one. We study the evolution of such bump cosmologies and their effects in the halo mass function and halo power spectrum using N-body simulations, the halo-model based \textsc{HMcode} method, and the peak background split framework. The bumps are subject to different nonlinear effects that become physically well understood, and from them we are able to predict that the most distinctive features will show up for intermediate halo masses $10^{12.3} \Mass < M < 10^{13.6} \Mass$. Out of this range, we expect halos are not significantly affected regardless of the location of the primordial bump in the matter power spectrum. Our analytical results are accurate and in very satisfactory agreement with the simulated data.  

\end{abstract}


\maketitle


\section{Introduction}

Recent cosmological and astrophysical observations have consolidated our picture of the concordance $\Lambda$CDM model \cite{Riess_1998,Perlmutter_1999,Planck2018b,2dFGRS:2005yhx,Eisenstein_2005}, which corresponds to a nearly homogeneous and isotropic expanding universe filled with the particles of the Standard Model (SM) \cite{ParticleDataGroup:2022pth}, and supplemented by dark matter and a cosmological constant. Despite this success, the two dark components have yet to be thoroughly tested and understood, since their fundamental nature is still a puzzle  \cite{Carroll_2001}. At present time, they accounts for about $96\%$ of the energy budget of the cosmos, and so alternative models look for plausible explanations. In particular, scalar fields have been proposed to describe dark energy \cite{Ratra:1987rm,Wetterich:1987fm,Caldwell:1997ii}, whose nature could be that of a fundamental particle not contained in the SM (Higgs-type particles) or can be a composite one, as for example a \textit{dark meson} pion $\pi$-like particle  \cite{delaMacorra:2018zbk}.

In addition to the fact that we do not know the nature of the dark sector, some strains in the $\Lambda$CDM model began to appear as the accuracy of cosmological observations improved, and recently, some interesting tensions have emerged. Perhaps the most famous is the discrepancy between early times and local measurements of the Hubble constant \cite{Bernal:2016gxb, Riess:2019cxk, Verde:2019ivm}. The increasing statistical tension on its value obtained using different observations has revived interest in alternative cosmological models. Henceforth, extensions to the standard model of particles have been proposed to alleviate the $H_0$ crisis or even simply to describe the origin of dark energy, for example by introducing additional particles.

As an example of the interest of this work, the Bound Dark Energy model (BDE) cosmological model \cite{delaMacorra:2018zbk, Almaraz:2018fhb} is characterised by a supersymmetric Dark Gauge Group (DG), in which the fundamental particles are massless during early times and their energy density evolved as radiation. However, at low energies the postulated gauge interaction becomes strong enough to bind the elementary dark particles together and form massive bound states, \textit{dark mesons} and \textit{dark baryons}. This process is similar to the strong QCD interaction in the SM where quarks are bound together to form baryons and mesons (e.g.~protons, neutrons or pions). In the BDE scenario, the dark energy corresponds to the lightest meson scalar particle $\phi$ formed at a phase transition scale $\Lambda_T$, at a scale factor $a_T$. 
Before the transition, the energy density of the DG particles behaves as radiation decaying with the expansion of the Universe as $1/a^4$. Just after the phase transition occurs, for a scale factor $a>a_T$ and lasting a long period of time, its energy density decays very fast, as $1/a^6$. During this epoch, there is an abrupt decrease in the DG cosmic abundance and rapidly becomes subdominant. We refer to such a behavior very generically as Rapid Diluted Energy Density (RDED). The existence of these type of bound particles modifies the evolution of the Hubble parameter $H$ and have also an important impact in the evolution of density perturbations, leaving distinctive signatures on cosmological distances, and in the matter power spectrum and other summary statistics around the corresponding transition scale $k_T=H(a_T)a_T$ \cite{delaMacorra:2018zbk,Almaraz:2018fhb,Gomez-Navarro:2020fef}. In particular, during the period where the extra energy density dilutes, the growth of matter fluctuations is enhanced. The order parameter is the Hubble constant $H(a)$ and the bump is related to a different expansion rate  noticeable in the ratio $H_\text{$\Lambda$CDMex}(a)/H_\text{$\Lambda$CDM}(a)$.

In this work we are motivated by the impact of a RDED in the matter power spectrum, which becomes enhanced around $k_T$ \cite{Gomez-Navarro:2020fef}. However, it is worth noting that the power spectrum and halo mass function can exhibit similar features due to the presence of various competing mechanisms. Several studies in the literature highlight the existence of similar features in different theoretical frameworks. For instance, the works \cite{Chantavat-Gordon-Silk-PRD2009,Erickcek:2011us,Erickcek-PRD-2015,Gosenca-Adamek-Byrnes-Hotchkiss-PRD-2017,delaMacorra:2018zbk,Redmond-Trezza-Erickcek-PRD-2018,Delos-Erickcek-Bailey-Alvarez-PRD-2018,Jaber-Bravo:2019nrk,Dante2020b,Bechtol:2022koa} among others, provide insights into the diversity of mechanisms that can give rise to distinctive features, as bumps, in the power spectrum and halo mass function. 
Given the wide range of theoretical models, we will work in a model-independent way with the introduction of a family of parameterized bumps into the linear matter power spectra $P_\text{$\Lambda$CDMex}$,\footnote{The suffix ``ex'' in the notational label ``$\text{$\Lambda$CDMex}$'' makes allusion to the extra energy density component not present in the $\Lambda$CDM model at early times, and referred below as $\rho_\text{ex}$.} which otherwise would be a $\Lambda$CDM one, where we will vary their positions and widths. Such a parametrization, as given in eq.~\eqref{linear-ps-bumpmodel} below, was proposed in \cite{Knebe:2001en}, and further used in \cite{Gomez-Navarro:2020fef}. By using these linear spectra as input, we will work beyond the linear regime using different complementary schemes, and focus on the consequences that the nonlinear evolution of the injected bumps has on halo clustering and halo abundance using the Peak-Background Split (PBS) framework \cite{Kaiser1984ApJ, Bardeen1986ApJ} and the Sheth-Tormen Halo Mass Function (HMF) \cite{Sheth:1999mn,Sheth-Mo-Tormen:2001}. Since our results exhibit some overlap with those presented in the paper by Knebe-Islam-Silk (KIS) \cite{Knebe:2001en}, we have dedicated an entire subsection (\S \ref{subsect:KIS}) in our study to comparing and contrasting our findings with theirs.

Although full N-body simulations successfully describe the nonlinearities of the matter clustering, they have the disadvantage of being computationally expensive.  Hence, in this work we use the approximated particle mesh N-body solver \textsc{FastPM}\footnote{\href{https://github.com/fastpm/fastpm}{https://github.com/fastpm/fastpm}} \cite{Feng:2016yqz}, where the linear growth of displacements, the Zeldovich approximation solution, is enforced at $k\rightarrow 0$ by choosing an appropriate set of kick and drift factors, and hence very large scale are treated exactly. Further, we use the \textsc{HMcode}\footnote{\href{https://github.com/alexander-mead/HMcode}{https://github.com/alexander-mead/HMcode}} \cite{Mead2015,Mead2016,MeadBrieden2020} to describe the nonlinear dark matter power spectrum. This is a halo-model based method, and as such, it describes the nonlinear power spectrum as a sum of two pieces, the 2-halo term that models the correlation between particles hosted by different halos reducing to the linear theory at large scales, and the 1-halo piece to model the small, intra-halo clustering scales \cite{Peacock:2000qk,Seljak:2000gq,Cooray:2002dia}.

Halo clustering is crucial in the study of the large-scale structure of the Universe, since it is governed by gravitational instability, responsible for the formation of dark matter halos and their distribution \cite{Wechsler:2018pic}. Furthermore, well physically motivated models often assume that galaxy formation is the result of the condensation of baryonic matter in already collapsed and virialized dark matter halos \cite{Lacey:1994su,Kauffmann:1993gv,Sheth:1999mn, Cooray:2002dia, Kravtsov_2012}. Therefore, the HMF is an important tool for studying the formation and evolution of galaxies.
The analytical understanding of these processes is also desirable, both to obtain a physical intuition, and for the study of different models and their wide range of parameters. In this work, we use the Sheth-Tormen HMF to describe the evolution of the number density of dark matter halos of a given mass and for our alternative cosmological models. However, since we are working beyond $\Lambda$CDM, we let free their parameters and fit them to data extracted from our N-body FastPM simulations. On the other hand, galaxy surveys show that at large scales, the number density fluctuation is roughly proportional to the matter density overdensity field, with a multiplicative factor called the bias  $b$ \cite{Kaiser:1984sw,Mo-White:1996}. We study this large-scale bias using PBS, which in addition to its utility, allows a physical interpretation of the halo bias \cite{Schmidt:2012ys}. 

Summarizing, we are interested in the distribution of nonlinear virialized objects because it allows us to check the evolution and final fate of the small primordial density fluctuations that have undergone gravitational collapse. Consequently, it is necessary to review the properties of halo statistics in models that have undergone a phase transition in early times. This the main topic of this work.

The rest of the paper is organized as follows. In Sec.~\ref{section:rded}, we briefly present how cosmic phase transitions lead to different cosmological signatures. In Sec.~\ref{section:modelling-materps}, we introduce the bump in the power spectrum $P_\text{$\Lambda$CDMex}$ and parameterize its width and position, in this section we also show the specifications of our N-body simulations suite. In Sec.~\ref{section:halo-abundance-clustering} we review the HMF formalism and the large-scale bias. We present our results and details for our bump cosmologies in Sec.~\ref{section:analysis-results}, with a detailed comparison to the work of \cite{Knebe:2001en}, and adding supplementary material in Appendix \ref{app:Nhalos}. Finally, in Sec.~\ref{section:conclusions} we present our conclusions. 

\begin{table*}
	\centering
	\begin{tabular}{l c c c c c} 
		\hline
		\\ [-3pt]
		Name & $N_{sims}$ & $\qquad A_T \qquad $ & $\qquad \sigma_T \qquad $ & $k_T \,\,[h \text{Mpc}^{-1} ]$ & $L_{box} \, \,[h^{-1}\text{Mpc} ]$ \\ [2pt]
		\hline
		\\ [-3pt]
		\textsc{medbump-k1} & 5 & $0.15$ & $0.3$ & $1.0$ & 1024\\[4pt]
		\textsc{thinbump-k1} & 5 & $0.15$ & $0.1$ & $1.0$ & 1024\\[4pt]
		\textsc{medbump-k0p5} & 5 & $0.15$ & $0.3$ & $0.5$ & 1024\\[4pt]
		\textsc{thinbump-k0p5} & 5 & $0.15$ & $0.1$ & $0.5$ & 1024\\[4pt]
		\textsc{$\Lambda$CDM} & 5 & $-$ & $-$ & $-$ & 1024\\[4pt]
		\hline
	\end{tabular}
	\caption{Specifications of our N-body simulation suite. 
	The background cosmological parameters are the same for all the simulations: $\Omega_m = 0.3$, $\Omega_b = 0.05$, $\Omega_{\Lambda} = 0.7$, $\Omega_{\nu} = 0$, $h = 0.7$, $n_s = 0.96$, $\sigma_8 = 0.8$. Each simulation uses $1024^3$ particles distributed over $N_{grid}=1024^3$ cells. We consider the redshifts $z=0,0.5,1,2$.}
	\label{tab:simulation}	
\end{table*}

\section{Rapidly Dilution Energy Density (RDED)}\label{section:rded}

Models beyond $\Lambda$CDM may leave different distinctive features in structure formation, as for example in the  matter power spectrum,  or cosmological distances \cite{Karwal:2016vyq,Poulin:2018dzj,delaMacorra:2018zbk,delaMacorra:2020zqv,Almaraz:2018fhb,Jaber-Bravo:2019nrk, Gomez-Navarro:2020fef,Erickcek-PRD-2015, Redmond-Trezza-Erickcek-PRD-2018}. As mentioned in the Introduction,  in this work we study models containing an extra energy density $\rho_{ex}$, beyond the standard $\Lambda$CDM model. At high energies, this extra fluid evolves as radiation $\rho_{ex}\propto 1/a^4$ with an equation of state (EoS) $w=1/3$. Later on, after a transition happens at a scale factor $a_T$, the  energy density evolves as  $\rho_{ex}\propto 1/a^6$ with an EoS $w=1$. 
Since the energy density dilutes rapidly after $a_T$, we refer to this effect as Rapid Diluted Energy Density  (RDED)  and has been studied  in  \cite{delaMacorra:2020zqv} and previously as a Steep Equation of State in \cite{Jaber:2017bpx,Jaber-Bravo:2019nrk,Devi:2019hhd}  and independently as ``Kination'' in \cite{Redmond-Trezza-Erickcek-PRD-2018} at a background and perturbation level.  
The presence of the extra fluid $\rho_{ex}$ at early times enhances the total energy density and the Hubble rate $H$,  decreasing the baryon-photon sound horizon, 
compared to the standard \LCDM model,  with the potential to ameliorate the Hubble tension \cite{Karwal:2016vyq,Poulin:2018dzj,delaMacorra:2021hoh,Ivanov:2020ril,2022arXiv221104492K}. At the same time the RDED of the extra fluid $\rho_{ex}$  leaves  detectable cosmological imprints as for example in cosmological distances  \cite{Jaber-Bravo:2019nrk} and in structure formation  \cite{Almaraz:2019zxy,Klypin:2020tud}.
%
Some of these models are generally referred as Early Dark Energy (EDE) in \cite{Linder_2008,Waizmann:2008zg,Francis:2008md,Grossi:2008xh} and more recently in \cite{Karwal:2016vyq,Poulin:2018dzj}.  The basic idea is to postulate an extra  component that contributes non negligible to the energy density at early times, and afterwards for  $a>a_T$ 
it dilutes faster than radiation with $w\simeq 1$. 

In the following, in order to describe the origin of a RDED we appeal to the mechanism of the Bound Dark Energy (BDE) model \cite{delaMacorra:2018zbk,Almaraz:2018fhb}. 
Following the BDE model, a plausible explanation of this RDED transition is if at high energies the original particles contained in $\rex$ are relativistic evolving as radiation but at a later time a phase transition takes place at $a_T$,  and  the original fundamental particles are bound together due to a strong interaction and a scalar potential $V(\phi)$  is generated, where $\phi$ corresponds to the lightest scalar composite field which corresponds to the bound dark energy (BDE) particle. The dynamics of BDE  particle produces the RDED evolution described above \cite{delaMacorra:2018zbk,Almaraz:2018fhb}.

Having this phase transition in $w$ affects not only the evolution of the energy density $\rex$, but the amplitude of matter perturbations is increased due to a RDED cosmological evolution  compared to a standard $\Lambda$CDM model, and a bump feature is produced in the ratio of power spectrum $P_\text{$\Lambda$CDMex}/P_\text{$\Lambda$CDM}$. Notice that this boost affects only matter fluctuations that entered the horizon before and around $a_T$, i.e. $k \geq k_T$. Modes entering during the rapid dilution (when $w=1$) are also affected, but only when the RDED has non-negligible impact on the expansion rate $H$. That is, as long as there is a noticeable change in $H$ due to the rapid dilution, the growth of perturbations will be impacted by RDED, and as more $\rho_{ex}$ dilutes (faster than radiation), the boost on the growth of matter becomes less prominent.  This is a characteristic signature on the matter power spectrum due to the presence of an early times RDED and was presented  in \cite{delaMacorra:2018zbk} and further developed in \cite{Almaraz:2018fhb,Jaber-Bravo:2019nrk,Gomez-Navarro:2020fef,delaMacorra:2020zqv}.

\section{Modelling the power spectrum}\label{section:modelling-materps}

In the following, we will characterize the effect of having a RDED by choosing a parametrization that we refer throughout as the bump cosmology, where the linear power spectrum is a modification to that of a standard $\Lambda$CDM cosmology one given by \cite{Knebe:2001en,Gomez-Navarro:2020fef}
\begin{equation} \label{linear-ps-bumpmodel}
P_\text{bump}(k, z)=\big[1+ F(k)\big]P_\text{$\Lambda$CDM}(k, z),
\end{equation}
with the $F(k)$ parametric function describing the bump,
\begin{equation} \label{eq:bumpmodel}
F(k)=A_T \exp \left(-\frac{[\ln(k/k_\mathrm{T})]^2}{\sigma_T^2} \right).
\end{equation}
The parameters $A_T$, $k_T$ and $\sigma_T$ are the amplitude, scale, and width of the bump, respectively. The width of the bump corresponds to how fast the rapid diluted energy density phase takes place, whereas $k_T$ represents the mode entering the horizon about the phase transition time.

We consider four different bump cosmologies, each with fixed amplitude $A_T=0.15$, as motivated by the BDE models.\footnote{In the original BDE model, the extra energy density has an abundance $\Omega_\mathrm{BDE}(a_T) \sim 0.11$ before the phase transition that occurs at $a_T = 2.48\times 10^{-6}$, and hence the mode entering the horizon at that time is $k_T \approx 0.92 \iMpc$ \cite{Almaraz:2018fhb}.} We choose two different widths of the bump $\sigma_T=0.3$, and $0.1$, and locate the bump at two different scales: $k_T=0.5$, and $1.0 \iMpc$ (see Table \ref{tab:simulation}). We consider these bump cosmologies at different redshifts: $z=0.0$, $0.5$, $1.0$, and $2.0$. 

Other different scales maybe worthy of study. For example, in the Early Matter Dominated scenarios only modes that entered the horizon very early in the universe, before Big Bang Nucleosynthesis, are enhanced \cite{Erickcek:2011us}. However such models presents features in the power spectrum at very small scales, completely out not of the reach of a cosmological evolution description. In the other hand, phase transitions can occur at late times in the matter dominated epoch, e.g. driven by a Steep Equation of State \cite{Jaber:2017bpx}, in such cases the bump is located at very large scales, and remain intact by the non-linear evolution. Therefore, we choose study bumps located at intermediate scales.

To generate the $\Lambda$CDM power spectrum we use the cosmological parameters reported in Sec.~\ref{subsect:NbodySims}, or in the caption of Table \ref{tab:simulation}, which are the same for all the bump and standard cosmologies, and as such, the only difference between the models is the presence of the bump parametrized by Eq.~\eqref{linear-ps-bumpmodel}.

In the following two subsections we briefly describe the methods we use to study the nonlinear evolution of the bump in the power spectrum.

\subsection{N-body simulations} \label{subsect:NbodySims}

We generate 25 rapid N-body simulations using the code \textsc{FastPM}, 5 for each of the cosmologies detailed in Table~\ref{tab:simulation}. Zeldovich initial conditions were generated at $z_i=99$, and we use $100$ linearly space steps up to redshift $z=0$. Each simulation uses $1024^3$ particles to approximate the density field. The box sizes of the simulations are $L_{box}=1024 \Mpc$. We analyze four snapshots at $z=0.0, 0.5, 1.0, 2.0$. 

Our baseline cosmology is $\Lambda$CDM with dark matter density $\Omega_{cdm}=0.25$, baryon density $\Omega_b=0.05$, fluctuation variance $\sigma_8=0.8$, dimensionless Hubble constant $h=0.7$, and spectral index $n_s=0.96$. Neutrinos are considered massless. 

We identify and construct halo catalogs with the Friends-of-Friends algorithm \cite{Davis:1985rj}, already implemented in the \nbodykit{} package\footnote{\href{http://nbodykit.readthedocs.io}{http://nbodykit.readthedocs.io}} \cite{Hand:2017pqn}, for which we use a linking length $l=0.2$ and where each halo is formed by at least 20 particles. 

\subsection{HMcode}

As a complementary tool to N-body simulations, we make use of the \textsc{HMcode} \cite{Mead2015,Mead:2016ybv}, which is an augmented version of the standard halo-model scheme for the nonlinear matter power spectrum. The starting point is a standard halo-model calculation, where the power spectrum is split into two terms: one that accounts for the clustering arising within individual halos, and the second that accounts for the clustering of dark matter between two different halos and that follows closely the linear theory at large scales. We use the most recent \textsc{HMcode} version \cite{MeadBrieden2020}, which further accounts for the BAO damping into the two-halo term. In this new scheme, \textsc{HMcode} adds a smoothing parameter for the transition region between the 1- and 2-halo terms when constructing full halo-model power spectra. In the nonlinear regime at low redshifts, we expect \textsc{HMcode} to match adequately the N-body simulations.

\section{Modelling the halo abundance and clustering}\label{section:halo-abundance-clustering}
Quantitative comparisons between theoretical predictions and observations allow us to compute constraints on cosmological parameters, e.g. the abundance of halo identified as clusters of galaxies. The bias of halo dark matter contains complementary information on their abundance since data survey is understood through the bias of the halos which they form \cite{Padilla-2dF-Galaxy-Survey:2004, Zehavi_2005}. In this section, we describe the analytical techniques to study the halo abundance and halo bias using the Sheth-Tormen mass function and PBS prescriptions.

\subsection{The halo mass function}\label{section:hmf}

The HMF gives the number density of dark matter halos as a function of their masses (per unit comoving volume). However, occasionally a good start point to the HMF discussion is by introducing the scaled differential mass function $f(\sigma)$ as the fraction of the total mass $\bar{\rho}V$ (when the volume $V$ is very large) hosted by halos in a logarithmic interval of $\sigma^{-1}$, 
\begin{equation}
    f(\sigma) = \frac{d (\rho / \bar{\rho})}{d \ln \sigma^{-1}},
\end{equation} 
with $\sigma(M)$ the variance of linear fluctuations when smoothed over a scale $R(M) = (3 M /4\pi \bar{\rho})^{1/3}$,
\begin{align}\label{eq:sigmaM}
\sigma^2 (M,z) = \frac{D^2(z)}{2\pi^2} \int_0^{\infty}  dk \, k^2 P_L(k) W^2(k R),
\end{align}
where $P_L$ is the linear power spectrum, $D$ the linear growth function and $W$ the top-hat filter in Fourier space
\begin{align}
    W(k R)= \frac{3}{(kR)^3}\big(\sin(kR)-kR\cos(kR) \big).
\end{align}
The utility of $f(\sigma)$ is that apparently this quantity, when properly scaled, is nearly universal throughout a wide range of halo masses, and for cosmologies both within the $\Lambda$CDM as well as dark energy models \cite{Bhattacharya:2010wy} or even in Modified Gravity theories \cite{Aviles:2018saf}.\footnote{This is more evident by comparing between different models the \textit{multiplicity} $\nu f(\nu)$ against the rescaled variance, or \textit{peak significance} $\nu=\delta_c/\sigma$ introduced below; see, e.g., Fig.~7 of Ref.~\cite{Aviles:2018saf}. Notice also that $\delta_c$ is mass dependent in theories that introduce new additional scales as in warm dark matter or modified gravity and hence comparing against $\nu$ and $\sigma^{-1}$ are not equivalent.} In contrast, the HMF is very sensitive to the cosmological parameters and the specific theory through the power spectrum dependence of the variance. 

The precise connection between $f(\sigma)$ and the HMF is given by
\begin{align}
\frac{dn}{d \log M} = f(\sigma) \frac{\bar{\rho}}{M}\frac{d \ln \sigma^{-1}}{d \log M},
\end{align}
where one assumes that all of the matter in the universe is hosted by halos.
The simplest HMF is the Press-Schechter mass function \cite{PressSchecter1974ApJ}, which analytical form can be obtained exactly based on the spherical collapse model and the hypothesis that the mass in collapsed objects is related to the volume with density above a certain threshold. However, data from simulations show that the Press-Schechter HMF is not accurate at the low and the high mass ends. Hence, several other mass functions have been proposed based on different assumptions or as purely empirical fits \cite{Jenkins:2000bv,Reed:2003sq,Warren:2005ey,Tinker:2008ff}. The Sheth-Torman HMF \cite{Sheth:1999mn,Sheth-Mo-Tormen:2001} is perhaps the best known and most used alternative to Press-Shechter, it is based on the more realistic ellipsoidal collapse and reproduce N-body simulations better. It is defined through 
\begin{align}\label{fST}
 f_{ST}(\nu) = A(p) \sqrt{\frac{2q}{\pi}}\left[1+\left(\frac{1}{q \nu^2} \right)^p \right] \exp \left(-\frac{q \nu^2}{2} \right),
\end{align}
with $A(p)=[1+\pi^{-1/2}2^{-p} \Gamma(0.5-p) ]^{-1}$ a normalization factor coming from the assumption that all the dark matter in the Universe is contained within halos. 
The variable $\nu= \delta_c / \sigma$ is the peak significance, and $\delta_c$ is the critical overdensity for collapse. Notice that we are abusing notation because $f(\sigma)$ and  $f(\nu)$ represent the same function. For a matter-dominated universe, $\delta_c=1.686$, while for $\Omega_m<1$ the numerical value varies slightly with the redshift but with no significant impact on the cosmological outputs. From Eq.~\ref{fST}, one recovers the Press-Schecter mass function by choosing $q=1.0$ and $p=0.0$. The standard values $p=0.3$ and $q=0.707$ for the Sheth-Tormen HMF are obtained by fitting to $\Lambda$CDM simulations. For the bump cosmologies, in Sect.~\ref{subsec:HMF} we will rely on Eq.~\eqref{fST} for computing the HMF, but we will fit the $p$ and $q$ parameters to our N-body simulations.

\subsection{Biasing on large scales}

To compute the halo bias we use the PBS formalism \cite{Kaiser1984ApJ, Bardeen1986ApJ, Sheth:1999mn, Sheth-Mo-Tormen:2001} with the aid of the Sheth-Tormen mass function. In this picture, long- and small-wavelength density fluctuations are split, with the crests of long wavelength overdensities serving as locations where the average density is higher than the background cosmological density, and on top of that small-wavelengths fluctuations, the peaks, collapse to form halos on which ultimately galaxy formation takes place. The main idea to obtain the halo bias is that more massive halos are formed in locations where the average (over large regions) local density is high, or in other  words the PBS biases are the responses of the mean abundance of tracers against small changes in the background density \cite{Sheth-Mo-Tormen:2001,Schmidt:2012ys}. That is, under the PBS formalism the Lagrangian ($L$) biases can be written in terms of the multiplicity function $\nu f(\nu)$ through
\begin{align}\label{PBSbLn}
b^L_{n}=\frac{(-1)^n}{\sigma^n(M,z)}\frac{1}{\nu f(\nu)} \frac{d^n \nu f(\nu)}{d \nu^n }.
\end{align}
It turned out that the bias computed in this way are the physical, renormalized local biases, up to subdominant factors introduced by the artificial smoothing scale of density fluctuations \cite{Matsubara:2008wx,Schmidt:2012ys,Aviles:2018thp}. 
For the Sheth-Tormen HMF given by Eq.~\eqref{fST}, the linear Lagrangian bias for halos of mass $M$ is 
\begin{align} \label{STb1}
    b^L_{1}(M)=\frac{1}{\delta_{c}}\left[ q \nu^2-1+\frac{2p}{1+(q\nu^2)^p} \right].
\end{align}
To obtain the biases over a mass range  $[M_{min}, M_{max}]$ one has to average over the above expression: that is, the large-scale bias within a mass range becomes \cite{Matsubara:2008wx}
\begin{align} 
    b_1^L = \frac{1}{I_{dM}}\int_{ M_{\text{min}} }^{M_{\text{max}}} dM \, \frac{1}{\delta_c} \frac{\nu}{M} \frac{\partial f}{\partial \nu} \frac{d \ln \nu}{dM},
    \label{eq:effective_bias}
\end{align}
with
\begin{align}
    I_{dM} = \int_{ M_{\text{min}} }^{M_{\text{max}}} dM \, \frac{f}{M} \frac{d \ln \nu}{dM}.
\end{align}
The corresponding linear Eulerian bias is given by
\begin{align}    \label{eq:pbs-bias}
    b_{1} = 1+b^L_{1}.
\end{align}
We will use these expressions when compute the halo biases for bump cosmologies in Sect.~\ref{subsect:hPSandBias}.

\begin{figure*}
\includegraphics[width=7 in]{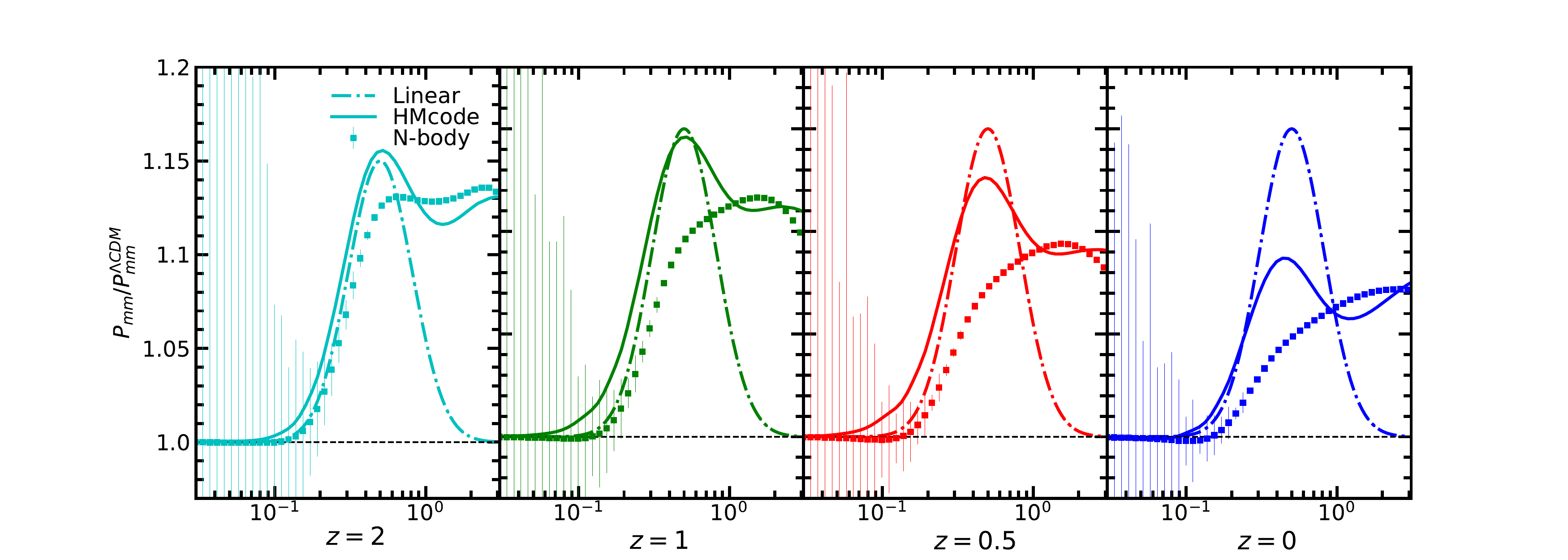}
\includegraphics[width=7 in]{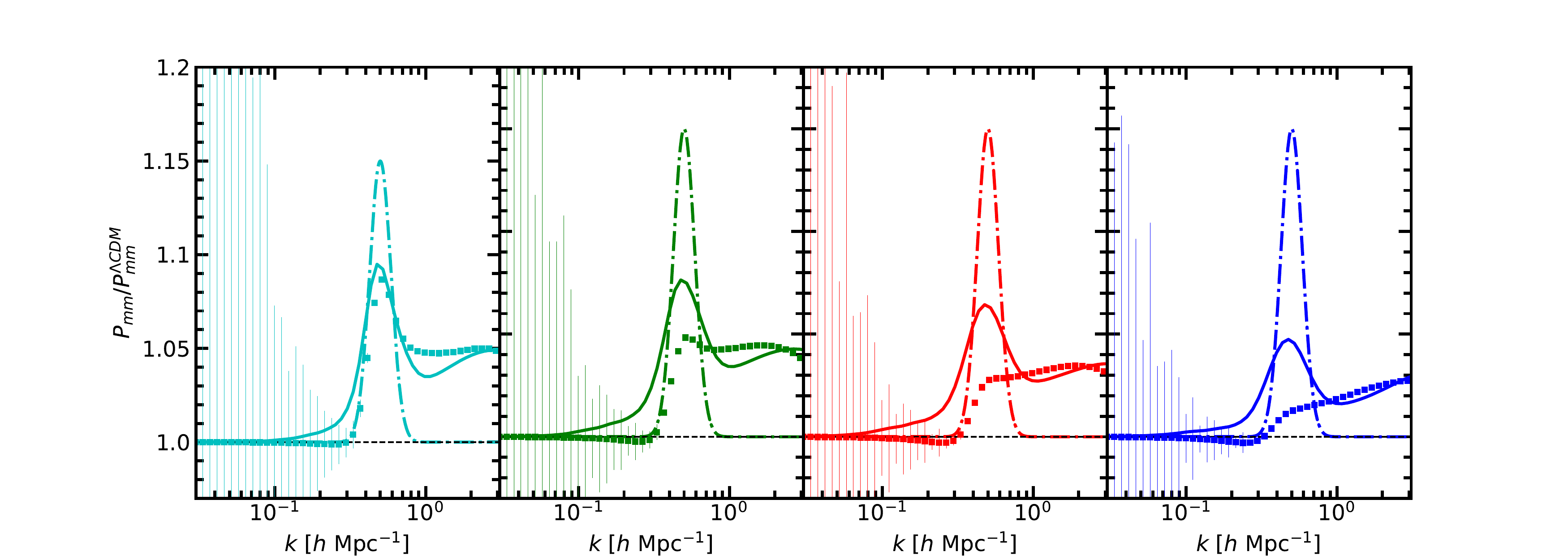}
\caption{Matter power spectrum response functions for bump cosmologies at $k_T=0.5 \iMpc$ with widths $\sigma_T=0.3$ (top panel) and $\sigma_T=0.1$ (bottom panel). From left to right, cyan curves are for $z=2$; green for $z=1$; red for $z=0.5$; and blue for $z=0$. Dot-dashed curves correspond to linear theory; solid to the \textsc{HMcode} model; and squares are the data extracted from the \fastpm N-body simulations. Error bars that are not visible are within the mark size. }
\label{fig:power-spectra-k0p5}
\end{figure*}

\begin{figure*}
\includegraphics[width=7 in]{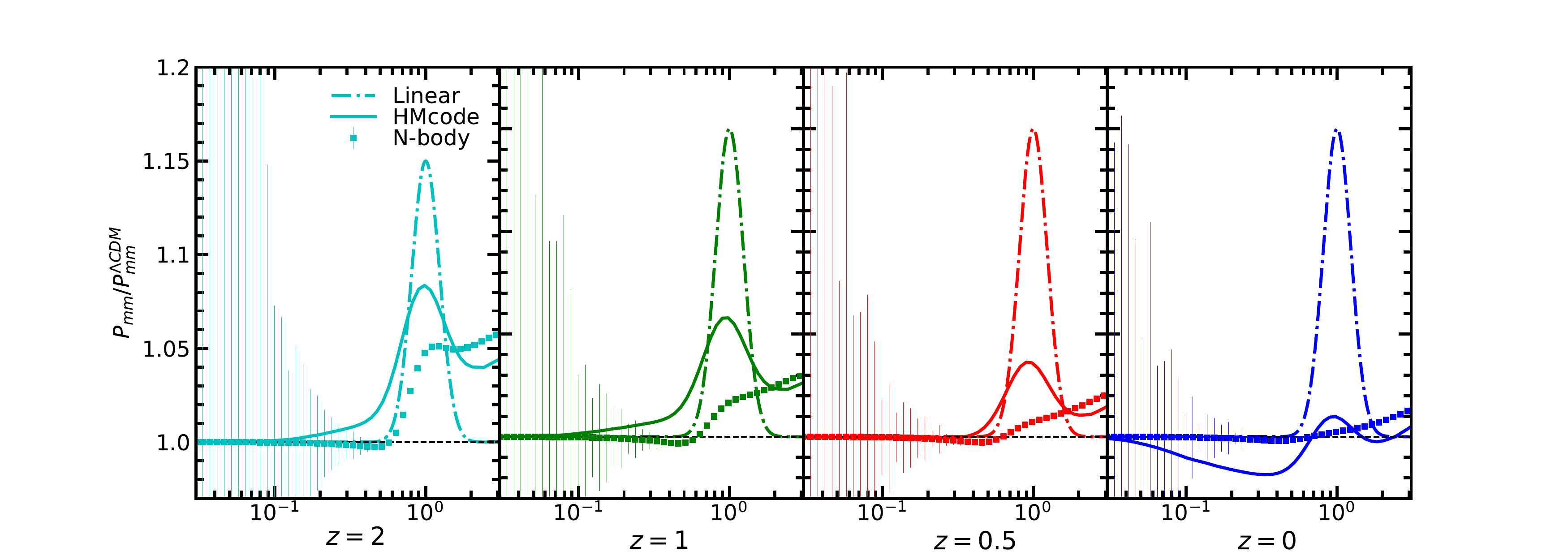}
\includegraphics[width=7 in]{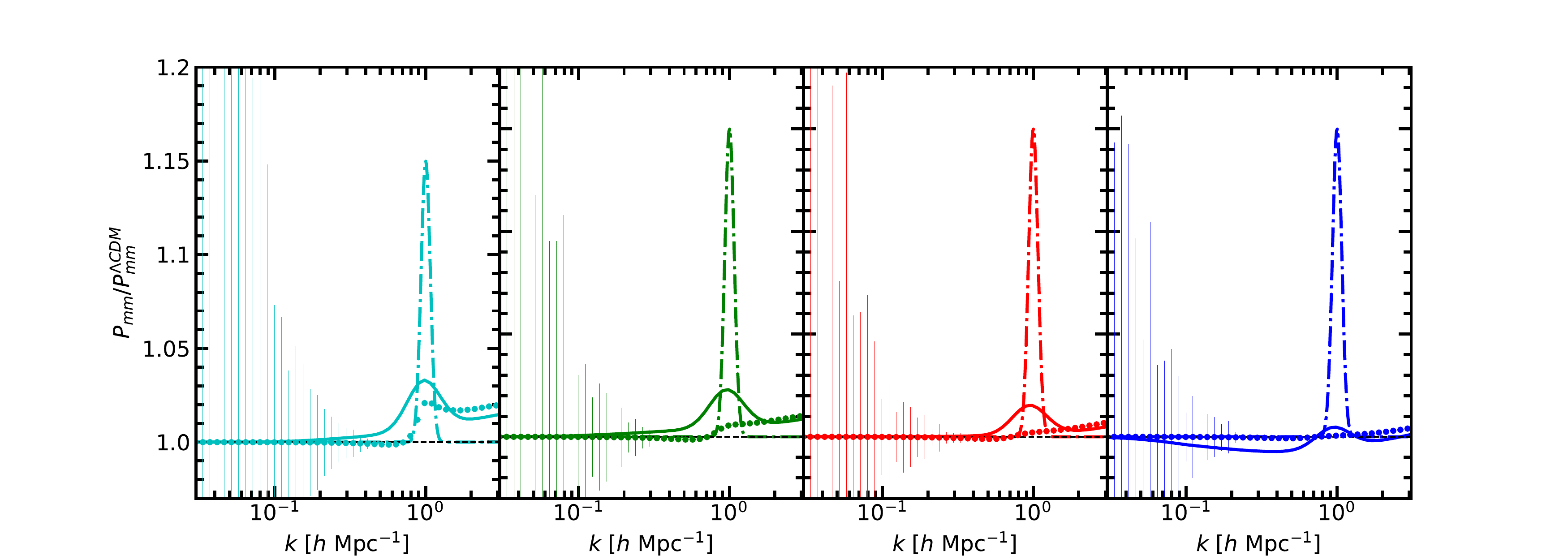}
\caption{Same as Fig.~\ref{fig:power-spectra-k0p5}, but for the bump cosmologies with $k_T=1 \iMpc$.}
\label{fig:power-spectra-k1}
\end{figure*}

\section{Analysis and results}\label{section:analysis-results}

The evolution, dilution and shift of the bump is studied via our suite of \texttt{FastPM} N-body simulations and the \textsc{HMcode} results. We first focus on response functions, computed as the ratio of statistics between a model containing a bump and one without it. For the power spectrum this is given by
\begin{equation} \label{responseR}
    R(k) = \frac{P_\text{bump}(k)}{P_\text{$\Lambda$CDM}(k)}.
\end{equation}
The importance of this analysis is that once a good understanding and modeling of the response function is acquired for a given alternative cosmology, its power spectrum can be computed by multiplying it by an as well modeled power spectrum for the $\Lambda$CDM one, which has been studied comprehensively. This response function analysis has shown to give fruitful results in different contexts beyond $\Lambda$CDM   \cite{Casarini:2016ysv,Mead:2016ybv,Cataneo:2018cic,Cataneo:2019fjp,Mead:2020qgo}.

Our numerical results are contrasted with the linear theory, for which the response in the power spectrum is simply $R_L(k) = 1+ F(k)$ at all $z$ since the linear growth is scale independent, well after the phase transition had occurred.

The details of our simulations are enlisted in Table \ref{tab:simulation}. The bumps are located at scales $k_T=0.5, 1.0 \iMpc$ and have widths $\sigma_T = 0.3$ and $0.1$. Hence, they are inside the full nonlinear regime, where a perturbative treatment is not adequate, as it is for the cases $k_T= 0.05$ and $0.1 \iMpc$ covered in our previous work \cite{Gomez-Navarro:2020fef}. Hence, in this work we use \textsc{HMcode} halo model and \fastpm simulations to model the nonlinearities, and do not rely in perturbation theory.
In \cite{Gomez-Navarro:2020fef} we also studied the case with transition scale $k_T= 1 \iMpc$, although only for the dark matter power spectrum and correlation function, while here we augmented the analysis for the case of biased tracers and put emphasis in the HMF and halo bias. When there is overlap with the above mentioned reference our results agree, despite that in that work we use full N-body simulations (but with a smaller number of particles: $N_p = 256^3$).

\subsection{Matter power spectrum}

To extract the power spectrum data we use the cloud-in-cell (CIC) mass-assignment scheme implemented in the \nbodykit{} package. The grid in our simulations is divided into $N_{grid}=2048$ cells and the size of the box in all cases is $L=1024\,h^{-1}\text{Mpc}$. The power spectrum ranges are binned in $80$ log-spaced $k$-points over the interval $[k_{min}, k_\mathrm{Ny}]$, where $k_{min}= 2\pi /L$ and $k_\mathrm{Ny} = N_{grid} \pi /L$ is the Nyquist frequency. Usually, power spectra in CIC approach are considered to be accurate up to half of the Nyquist frequency, and as such, this is the upper limit we show in our plots.

In Figs.~\ref{fig:power-spectra-k0p5} and \ref{fig:power-spectra-k1} we show the matter power spectra using the bump cosmology located at the scales $k_T=0.5$ and $1\iMpc$, respectively. The matter power spectra are computed using our different nonlinear methods and then divided by their counterparts in the $\Lambda$CDM model to show the response function; see Eq.~\eqref{responseR}. These analyses compare how the bump cosmology power spectra are modified by nonlinearities within the different schemes. The squares correspond to the \fastpm synthetic data and the error bars are the scattering over the 5 simulated boxes for each model.

As expected, at higher redshifts nonlinear effects are smaller and the responses for all approaches are very similar. The observed features in the plots can be described through two important nonlinear effects
\begin{itemize}
\item [1.-] There exists a generation of a second bump. This nonlinear effect was observed first in \cite{Gomez-Navarro:2020fef} and is due to that the primordial bump enhances the amplitude of the long wavelength perturbation where peaks in the density fluctuation locate, which magnify them, and ultimately forming a second bump. In other, simpler words, the generation of the second bump is a consequence of the development of structures on top of the first, primordial bump. This effect is even more pronounced for wider bumps because these provide a greater enhancement of linear power and then, in the language of halo-based models, a broader range of interaction with the 1-halo term, as explained in Ref.~\cite{Gomez-Navarro:2020fef} within the context of the halo model.  For the case of $k_T=0.5 \iMpc$, as can be observed in Fig.~\ref{fig:power-spectra-k0p5}, this nonlinear second bump is well modelled by the \textsc{HMcode}, and reaches a maximum relative to the first bump value,  at $z=0$.  For $k_T= 1 \iMpc$ the generation of this second bump is still present, although less evident because the Nyquist frequency coincides with the onset of the bump; see Fig.~\ref{fig:power-spectra-k1}.   

\item [2.-]  The primordial bumps tend to vanish with the gravitational collapse. This effect can be better understood under a configuration space description, where localized features become oscillations. Since bulk  displacements of matter tend to partially move out from overdense regions and populate underdense regions, these oscillations are erased with the collapse. In fact, the characteristic scale for this to happen is given by two times the Lagrangian displacements variance
\begin{align}
2 \sigma_\Psi &= 2 D(z) \left[\int_0^\infty \frac{dk}{6 \pi^2} P_L(k,z=0) \right]^{1/2} \nonumber\\
            &\sim  10 D(z) \,h^{-1}\text{Mpc},
\end{align}
with $D(z)$ the linear growth function. As can be seen by comparing Figs.~\ref{fig:power-spectra-k0p5} and \ref{fig:power-spectra-k1}, as higher is the transition mode $k_T$, the more the bump is damped. This is expected because higher-$k$ locations of the bump in Fourier space correspond to  higher frequency  of oscillations in configuration space, and since particles travel in random paths an average distance $ \sigma_\Psi$, then with high probability, particles will tend to displace out from the overdense regions if the oscillations wavelength are smaller. Of course this is a consequence of having a comoving distance between peaks and troughs in the 2-point configuration space correlation function comparable or smaller than $2 \times \sigma_\Psi$. So, it is not expected to happen for bumps located at low $k$ values, as those studied in \cite{Gomez-Navarro:2020fef}, which oscillations in configuration space have very large wavelengths. Notice this effect is completely nonlinear in the matter overdensity, i.e. it is non-perturbative in the Eulerian theory, however, it can be completely described by Lagrangian perturbation theory, even at the linear order, Zeldovich approximation because bulk, large scale displacement fields are responsible for erasing it, as it is shown in \cite{Gomez-Navarro:2020fef} (see particularly Sect.~4.3 and Fig.~8, where Convolution-LPT \cite{Carlson:2012bu,Vlah:2015sea} and simulated data are in very good agreement). Actually, we notice this effect has the same origin as the smearing of the BAO peak.
\end{itemize}

\begin{table*}
	\centering
	\begin{tabular}{l c c } 
		\hline
		\\ [-3pt]
		Name & $\qquad q \qquad $ & $\qquad p \qquad $  \\ [2pt]
		\hline
		\\ [-3pt]
		\textsc{medbump-k1}  & $0.727$ & $0.314$ \\[4pt]
		\textsc{thinbump-k1}  & $0.734$ & $0.326$ \\[4pt]
		\textsc{medbump-k0p5}  & $0.723$ & $0.343$ \\[4pt]
		\textsc{thinbump-k0p5}  & $0.691$ & $0.334$ \\[4pt]
		\textsc{$\Lambda$CDM}  & $0.711$ & $0.317$  \\[4pt]
		\hline
	\end{tabular}
	\caption{Best-fit for $p$ and $q$ Sheth-Tormen parameters. These results were obtained by using the criterion that minimizes Eq.~\ref{eq:minimize-quantity} at redshift $z = 0$. }
	\label{tab:STmassfunction}	
\end{table*}

\begin{figure*}
\includegraphics[trim={2cm 0.6cm 2cm 1cm}, height = 0.450\textheight]{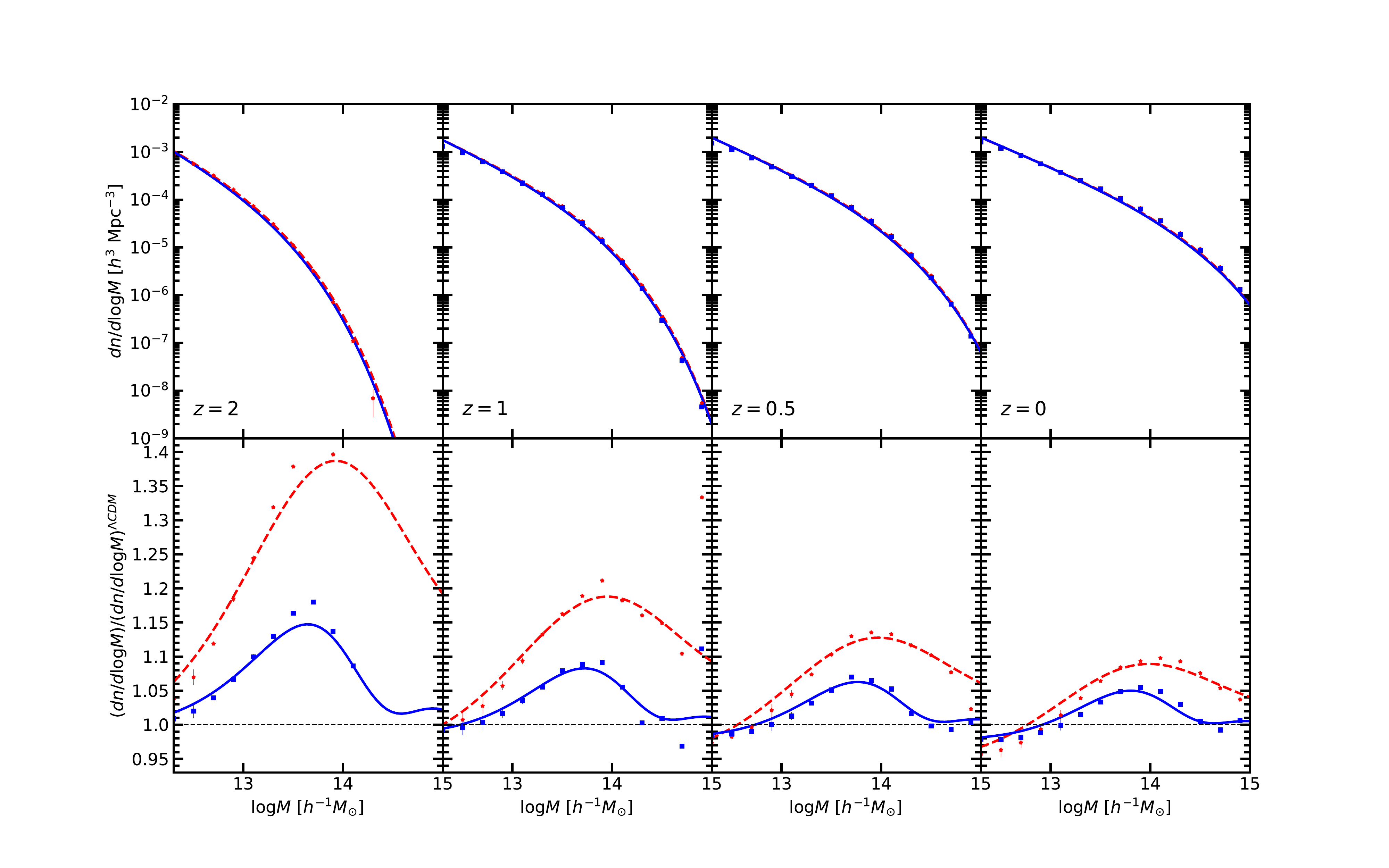}
\includegraphics[trim={2cm 0.6cm 2cm 1cm}, height = 0.450\textheight]{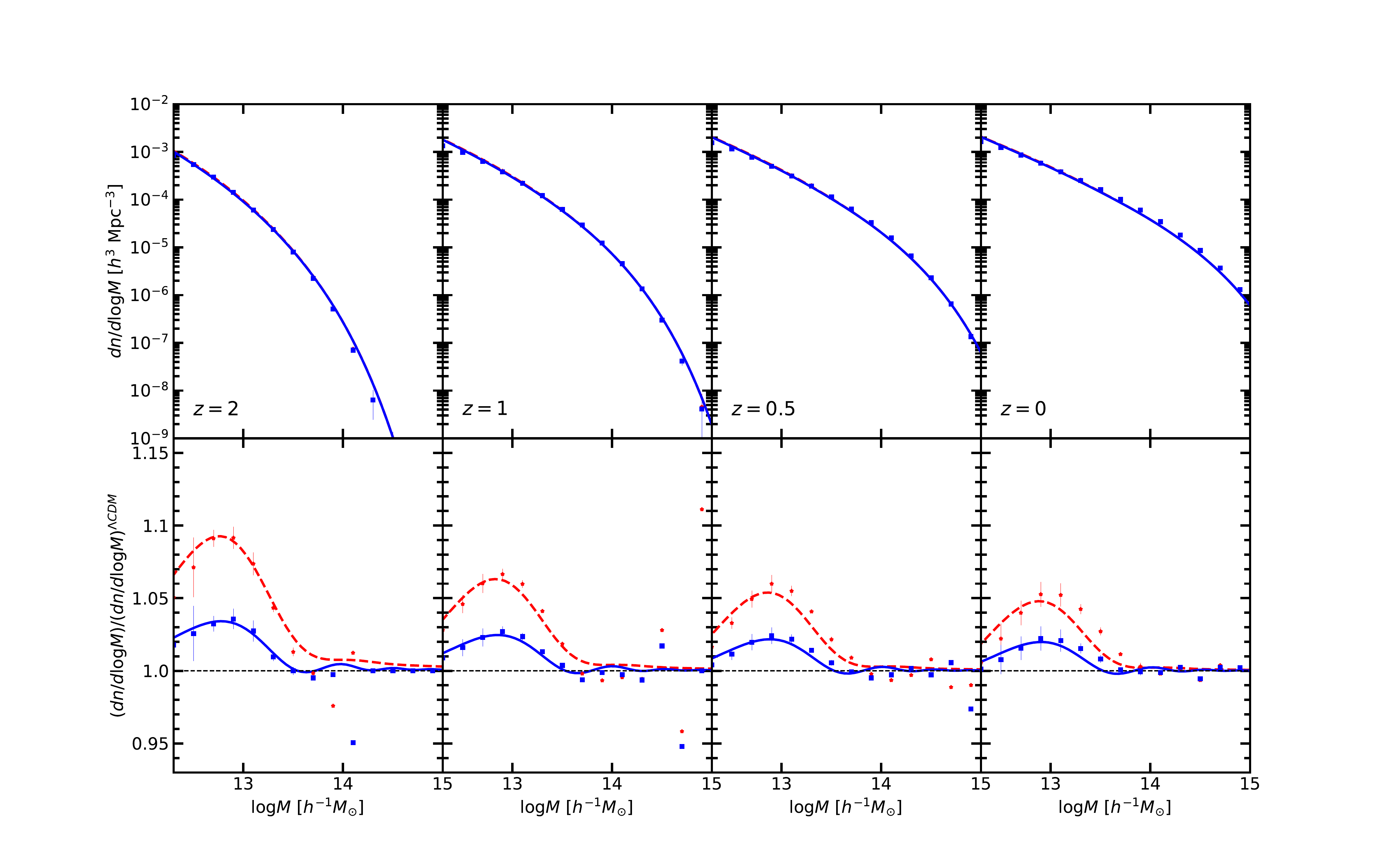}
\caption{Halo mass function $dn/d\log M$ for the bump cosmologies at $k_T=0.5 \iMpc$ (top panel) and $k_T=1 \iMpc$ (bottom panel). The dashed curves are for the Sheth-Tormen mass function of the bump cosmologies with $\sigma_T=0.3$; solid for the widths $\sigma_T=0.1$; stars are for the measurement from N-body simulations with $\sigma_T=0.3$; and squares for $\sigma_T=0.1$. Error bars that are not visible are within the mark size.}
\label{fig:hmf-k0p5-k1}
\end{figure*}

\subsection{Halo mass function} \label{subsec:HMF}

From our N-body simulations, we have constructed halo catalogs with the Friends-of-Friends algorithm as described in Sec.~\ref{subsect:NbodySims}. From them, we obtain the HMF for the different bump cosmologies. 
The number of halos per logarithmic mass intervals $dn/d \log M$ in a simulation of volume $L^3$ is given by
\begin{align}
\frac{dn}{d \log M} = \frac{M}{L^3 } \frac{\Delta N}{\Delta \log M},
\end{align}
for which we construct logarithmic bins in mass with size $\Delta \log M =0.2$. Complementary, in Appendix \ref{app:Nhalos} we show a histogram of our halos within the interval $M= [10^{12.3},M=10^{15}] \Mass$ as well as a Table with the mean number of halos of our catalog suite. 

Since we are working in cosmologies beyond the $\Lambda$CDM, and anticipating possible deviations in the scaled differential mass function $f(\sigma)$ function,  we recalibrate the model parameters of the Sheth-Tormen functional form of Eq.~\eqref{fST}. That is, we compute the best fit of Sheth-Tormen $p$ and $q$ parameters using the criterion that minimizes the quantity
\begin{align}
    \sum_i \left | \frac{n_{sims}(M_i)}{n_{model}(M_i,p,q)}-1 \right|.
    \label{eq:minimize-quantity}
\end{align}
where the sum runs over all the $\Delta \log M$ intervals. To perform the minimization we used the set of halo counts at redshift $z=0.0$. The best fit values are detailed in Table \ref{tab:STmassfunction}. The results seem to be consistent with the Sheth-Tormen model, since the deviations of the best fits for all the different models, including LCDM, with the values $q = 0.707$ and  $p = 0.3$ are quite similar. This analysis shows the universality of the Sheth-Tormen HMF, or more precisely of the scaled differential mass function, or multiplicity $f(\sigma)$, in these scenarios. Also for this reason, when we use the \textsc{HMcode} we do not change the Sheth-Tormen parameters.

Even though, we do expect the HMF to be sensitive to the nature of dark energy for bump cosmologies since the dependence of the variance $\sigma$ on the mass is different to that of $\Lambda$CDM. 
We anticipate more pronounced signatures in halos of intermediate masses, which we will explain below.
In Fig.~\ref{fig:hmf-k0p5-k1} we show the HMFs measured from the simulated data, as well as the ratio between them in the bump and $\Lambda$CDM cosmologies, both at $k_T=0.5$ (top panel) and $k_T=1.0 \iMpc$ (bottom panel). We do this for different redshifts, from left to right, these are $z=2.0, 1.0, 0.5, 0.0$. It is evident and expected that the largest deviations from the standard model occurs in the \textsc{medbump} cosmologies ($\sigma_T=0.3$) since these provide the largest enhancement of the linear power spectrum at the bump location. 

The upper panel of Fig.~\ref{fig:hmf-k0p5-k1} shows the HMF for the transition mode $k_T=0.5 \iMpc$. The largest deviation from $\Lambda$CDM occurs for halos with intermediate mass $M \sim [10^{13.4} , 10^{13.9}] \Mass$, reaching a difference of $38 \%$ ($10 \%$) at $z=2.0$ ($z=0.0$) for \textsc{medbump-k0p5} cosmologies. At the lowest redshifts snapshots ($z=0.5,0.0$), there are more small halos in the $\Lambda$CDM model than in the bump cosmologies. On the other hand, for the \textsc{thinbump-k0p5} cosmology we see that for very massive halos, the HMF is the same as that in $\Lambda$CDM at late times. We notice from the panels in Fig.~\ref{fig:hmf-k0p5-k1} showing the absolute HMF $d n/d\log M$, instead of its ratio to the $\Lambda$CDM one, that the number of halos per mass interval increases with redshift for both $\Lambda$CDM and bump cosmological models, which means that massive structures are been formed from these initial density fields. Nevertheless, the differences between the HMFs diminish with time. This can be expected because the nonlinear evolution tends to erase the bumps in  the matter power spectrum, as explained in detail in the previous section, and this naively could be translated into a expectation of a reduction on the excess of massive halos in bump cosmologies. However, the exact mechanism behind this reduction is not entirely clear to us, as virialized structures are unlikely to be destroyed by large-scale bulk flows of matter. An alternative explanation could be that the attenuation of the bump in the HMF results from a higher rate of mergers occurring at those mass scales, as proposed in \cite{Erickcek-PRD-2015}.

For the bumps located at $k_T=1.0 \iMpc$, qualitatively similar results are shown in the bottom panel of Fig.~\ref{fig:hmf-k0p5-k1}, but now the differences with $\Lambda$CDM are located at smaller halo masses $M \sim [10^{12} , 10^{13}] \Mass$. The HMF of the \textsc{medbump-k1} cosmologies (\textsc{thinbump-k1}) reaches a difference with respect of $\Lambda$CDM of $\sim 9 \%$ ($\sim 3 \%$) for halos of mass $12.5 \Mass$, which decreases at late times for the same reasons explained above in the case of the transition scale $k_T=0.5 \iMpc$. These differences between \textsc{medbump-k1} cosmologies and $\Lambda$CDM fall below the $1$ per cent for massive halos with mass $M > 10^{14} \Mass$.

The natural mass scale for the appearance of the bumps in the HMF is settled by the relation $k_T \simeq 2\pi/(4 R)$, where $R(M)$ is the Lagrangian radius of the halo. This gives $M(k_T) \simeq 4.5 \times 10^{12} \,\Omega_m (k_T /( h\, \text{Mpc}^{-1}))^{-3}  \, \Mass  $, which yields on average 5 times smaller than the mass values at which we found the bumps.

The above mentioned departures from $\Lambda$CDM are purely a consequence of the RDED in the bump cosmologies. These are the consequence in the increase of the halo formation relative to a cosmology with no bump, because the HMF is related to the standard deviation in the density field when smoothed over the Lagrangian radius $R(M)$. When the enhancement of the matter fluctuations appears a higher $k$, it affects smaller scales and then smaller massive halos are expected to be more abundant than in $\Lambda$CDM. On the other hand we have seen that as higher is the transition mode $k_T$, the bump is more susceptible to be erased by the nonlinear evolution, and hence the above picture cannot continue indefinitely and on the lower mass tail of the HMF range we always expect to obtain back the $\Lambda$CDM results.  That is, we expect to be able to test these kinds of cosmologies with intermediate massive halos, being these the most useful to constrain dark energy models. This can be put in contrast to alternative dark matter models, where the largest differences from CDM are expected to show up for the lowest massive halos, as e.g. in warm or fuzzy dark matter models in which the abundance of very low massive halos is suppressed, if they can be formed at all \cite{Lovell:2013ola,LinaresCedeno:2020dte,Kulkarni:2020pnb}.

Based on the preceding section discussion, one would naturally expect that a second peak in the matter power spectrum would correspond to a second peak in the HMF. However, this second bump is not observed in Fig.~\ref{tab:STmassfunction}, with the exception, perhaps, of a small deviation in the Sheth-Tormen theoretical curves with a narrow width of $\sigma_T = 0.1$. It is important to note that the second bump in the matter power spectrum did not arise during the formation of protohalos at high redshifts. Instead, these peaks emerged due to nonlinear collapse later on. Consequently, if a second peak were to appear in the HMF, it would be significantly suppressed. Additionally, it should be considered that our simulated halos have limitations, and the statistical significance is reduced for very massive halos. Specifically, in Figure ~\ref{tab:STmassfunction}, no discernible pattern is observed in the ratio of the HMF for bump cosmologies and the $\Lambda$CDM model above $10^{14} \Mass$.

\begin{figure*}
\includegraphics[width=7 in] {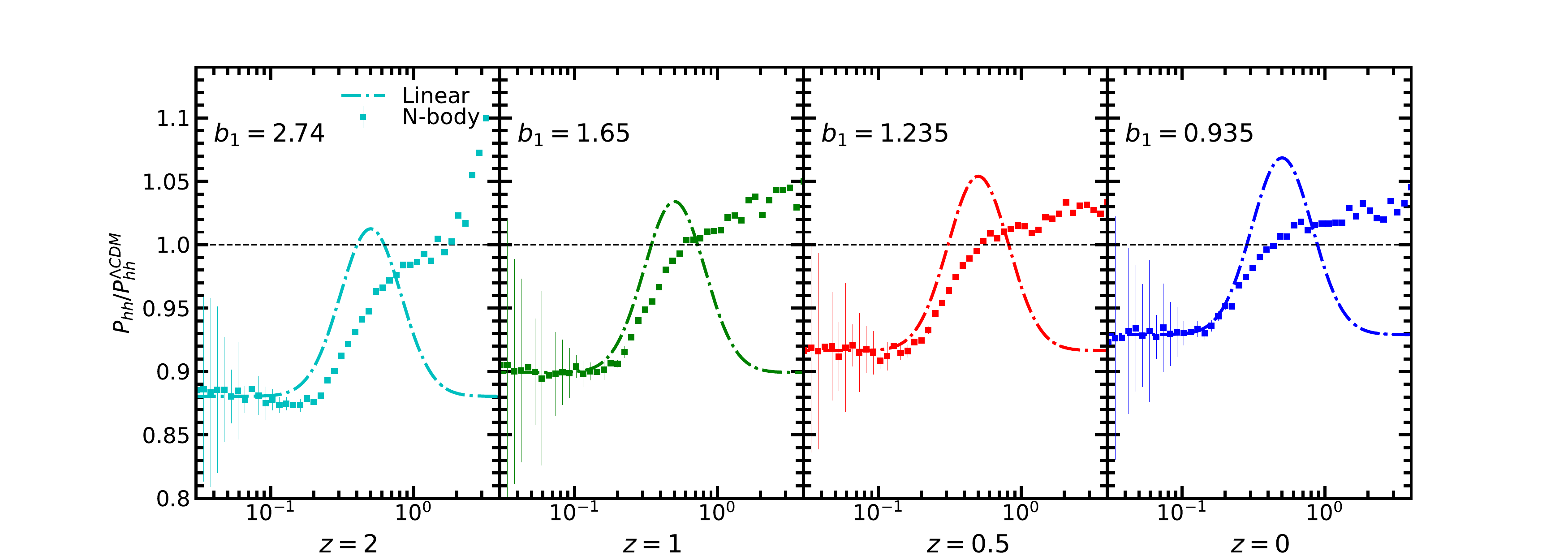}
\includegraphics[width=7 in] {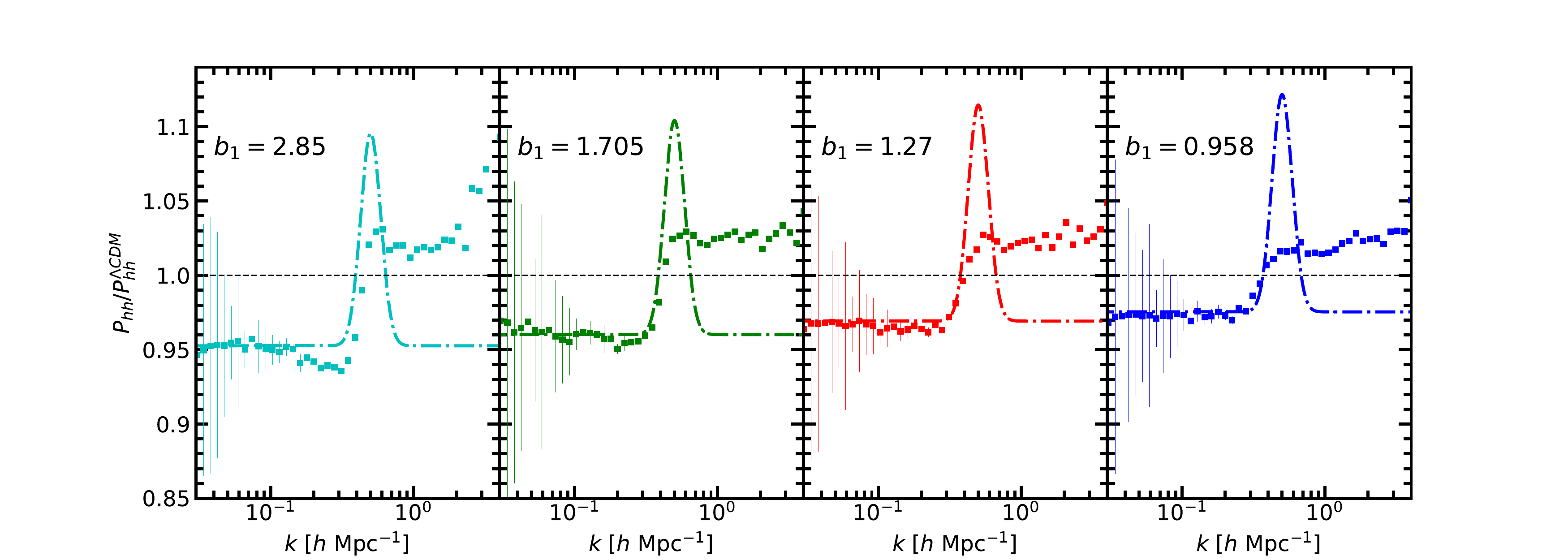}
\caption{Halo power spectrum for the bump cosmologies at $k_T=0.5 \iMpc$ for $\sigma_T=0.3$ (top panel) and $\sigma_T=0.1$ (bottom panel). We consider halos in the mass interval $[10^{12.3},10^{13}] \Mass$. From left to right, cyan curves are for redshift $z=2$; green for $z=1$; red for $z=0.5$; and blue for $z=0$. The dot-dashed curve corresponds to the linear theory; and squares are for the measurement from N-body simulations. Error bars that are not visible are within the mark size.}
\label{fig:halo-power-k0p5}
\end{figure*}

\begin{figure*}
\includegraphics[width=7 in] {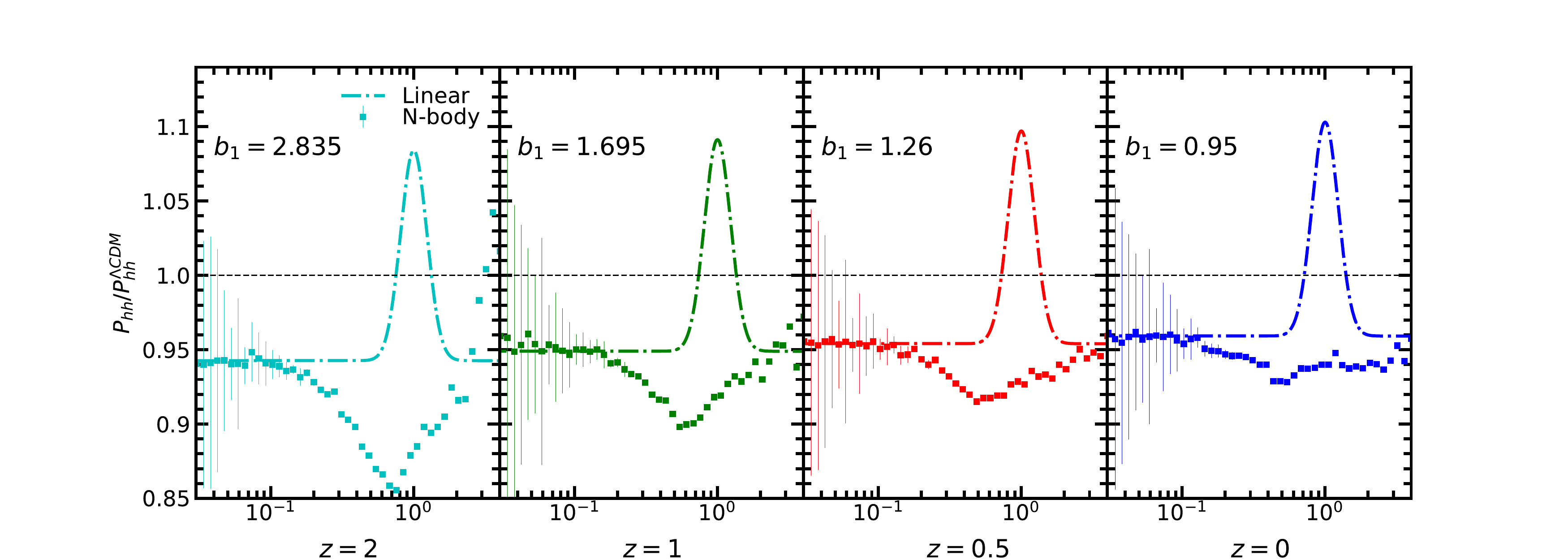}
\includegraphics[width=7 in] {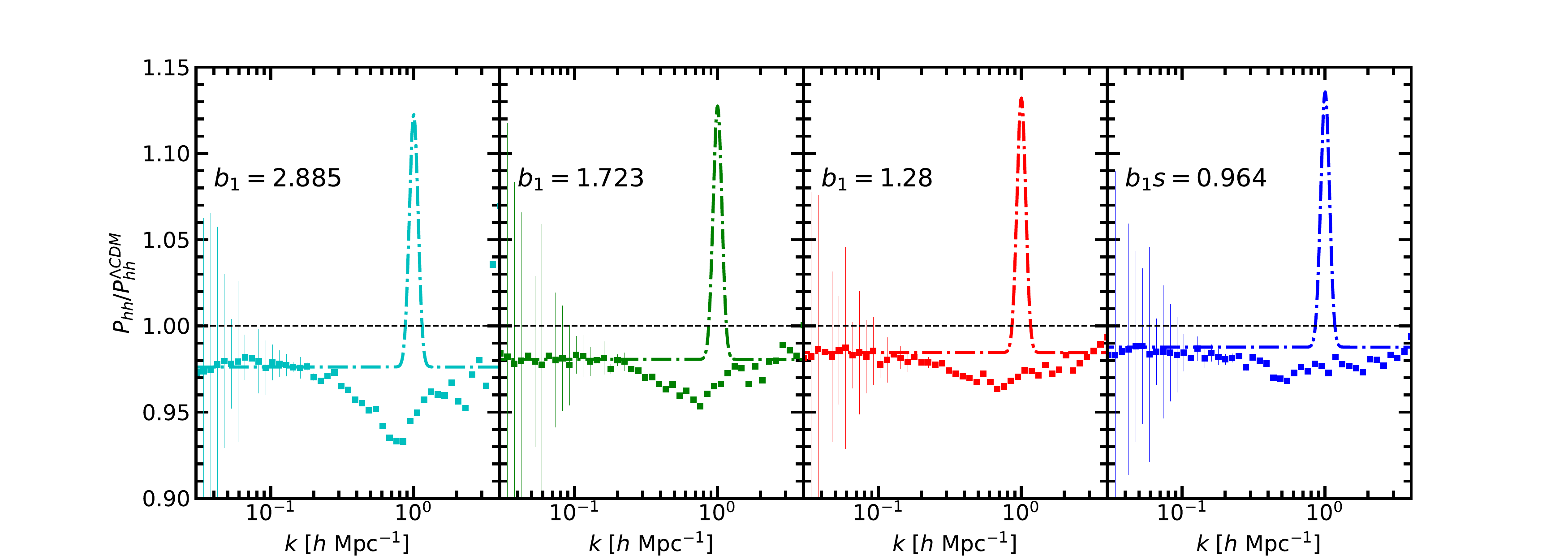}
\caption{Same as Fig. \ref{fig:halo-power-k0p5} but for the bump cosmologies with $k_T=1 \iMpc$.}
\label{fig:halo-power-k1}
\end{figure*}

\subsection{Halo power spectrum and large scale bias} \label{subsect:hPSandBias}

Now, we show results for the clustering of dark matter halos in Fourier space. For this analysis we first focus in the mass interval $10^{12.3} \Mass  <M<10^{13.0} \Mass$. 
%

In Figs.~\ref{fig:halo-power-k0p5} and \ref{fig:halo-power-k1} we show the halo power spectrum for bump cosmologies located at $k_T=0.5 \iMpc$ and  at $k_T=1 \iMpc$, respectively, with $\sigma_T=0.3 \iMpc$ (top panels) and $\sigma_T=0.1 \iMpc$ (bottom panels). On each panel, we show the response function given by the ratio of the halo power spectra in bump and $\Lambda$CDM cosmologies; see Eq.~\eqref{responseR}. We notice a weaker clustering at large scales in the bump cosmologies compared to $\Lambda$CDM. However, for small scales, for modes that entered the horizon before the transition, the situation changes drastically and the clustering is enhanced in the bump cosmology. The latter effect is expected because small scales tend to grow larger since they lie on top of modes around $k_T$ that were affected by the phase transition of the RDED. However, this only happens in the case of $k_T= 0.5 \iMpc$, 
while for $k_T=1.0 \iMpc$ the bumps are completely erased, as can be seen from Fig.~\ref{fig:halo-power-k1}. We believe this result deserves further investigation. Nevertheless, we also notice that the corresponding bump in the matter power spectrum of Fig.~\ref{fig:power-spectra-k1} is considerably more suppressed in comparison to the case of $k_T= 0.5 \iMpc$ shown in Fig.~\ref{fig:power-spectra-k0p5}. 

\begin{figure*}
        \includegraphics[width=7 in]{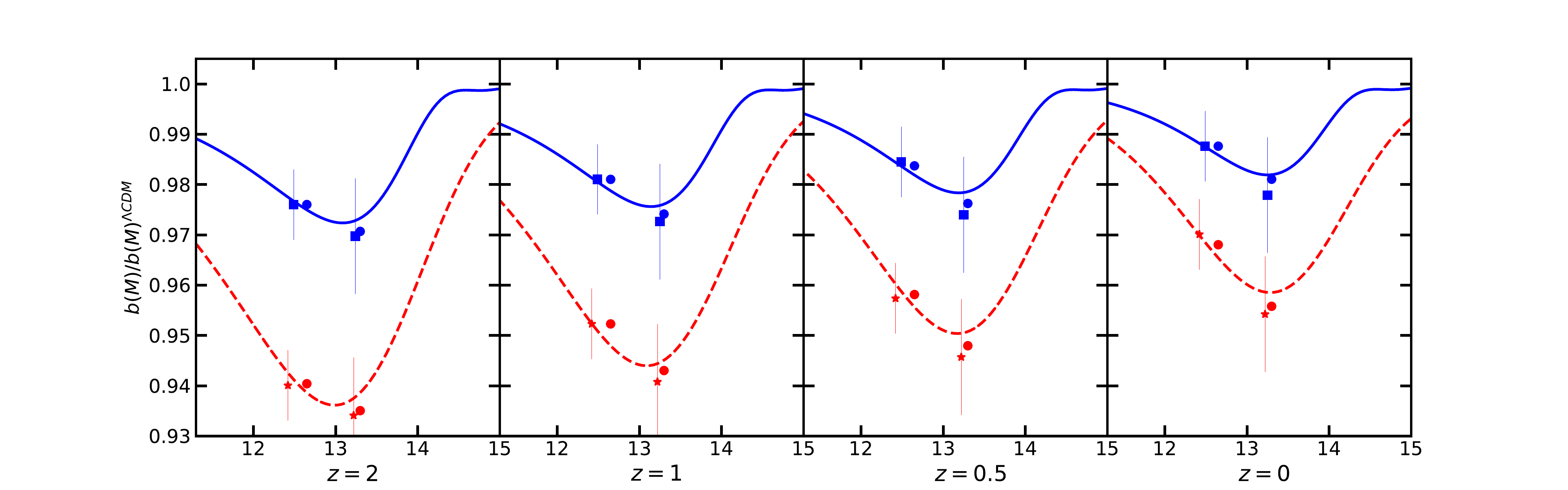}
    \includegraphics[width=7 in]{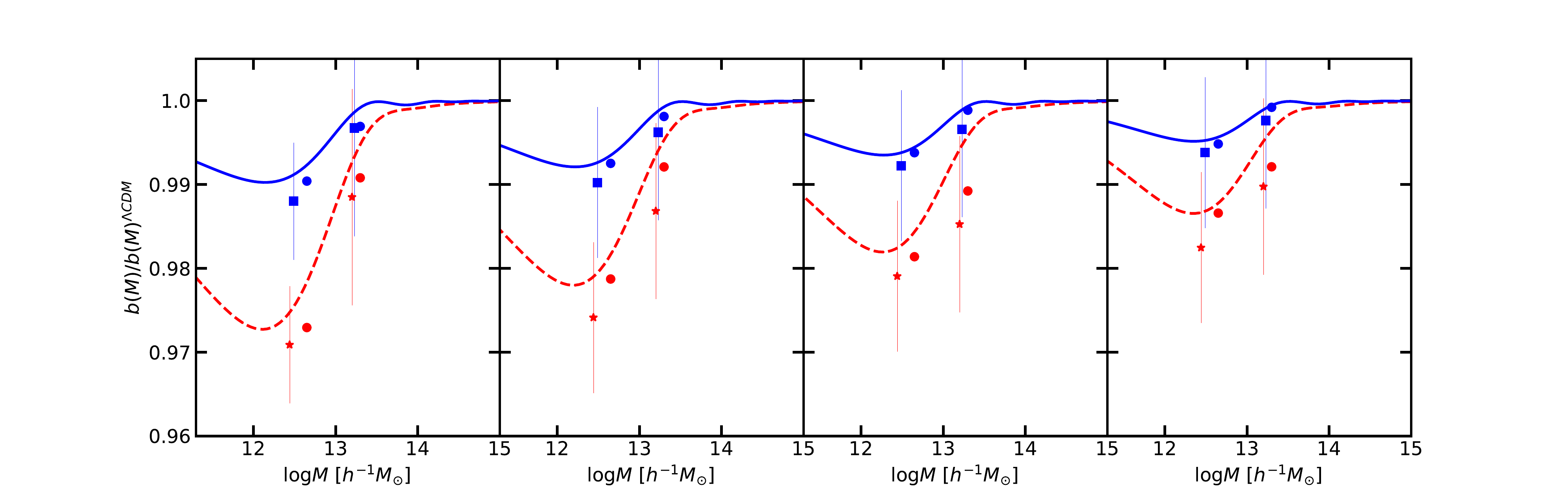}
\caption{Comparison between the halo bias from N-body simulations (star and square markers) and from theoretical prediction using PBS approach (solid and dot-dashed curves) for bump cosmologies at $k_T=0.5 \iMpc$ (top panel) and at $k_T=1.0 \iMpc$ (bottom panel). Stars are for the bump cosmologies with $\sigma_T=0.3$; and squares for $\sigma_T=0.1$. Dashed curves are for $\sigma_T=0.3$; and solid are for $\sigma_T=0.1$. Circles are for the effective bias defined in Eq. (\ref{eq:effective_bias}) in the mean ranges of $[10^{12.3} \Mass, 10^{13} \Mass]$ and $[10^{13} \Mass, 10^{13.6} \Mass]$.}
\label{fig:halo-bias-k0p5-k1}
\end{figure*}

The fact that at large scales one obtains a weaker clustering is less intuitive, and as we see below, is due to a different mass function and hence a different large scale bias. 
Indeed, as a consequence of the biasing being different between the different models, the response is not equal to unity at large scales. To understand this, in Fig.~\ref{fig:halo-bias-k0p5-k1} (solid lines) we show the theoretical dark matter halo bias as function of halo mass $b_1(M)$, as obtained from the PBS formalism and the Sheth-Tormen HMF using Eq.~\ref{STb1}. We do this for the bumps located at $k_T=0.5 \iMpc$ (top panel) and $k_T=1.0 \iMpc$ (bottom panel), and we plot their ratios to the $\Lambda$CDM biases. This analysis shows that the large scale offsets (departing from unity) observed in Figs.~\ref{fig:halo-power-k0p5} and \ref{fig:halo-power-k1} are due to different HMFs, and that the computed values with the PBS recipe match very accurate the simulations. In these figures,  the used linear power spectra are given by $b_1^2 P_L(k)$ with the linear bias $b_1$  computed  theoretically using Eq.~\eqref{eq:effective_bias}.  The differences in the biases are mainly  a consequence of the dependence of the variance of fluctuations with the halo mass, $\sigma(M)$, and not to the difference in the obtained $p$ and $q$ parameters for each models, which is actually small, certainly not enough to explain this large discrepancy on the biases.

Together with the theoretical results for $b(M)$, in Fig.~\ref{fig:halo-bias-k0p5-k1} we show the bias estimated by taking the ratio between the simulated nonlinear and linear power spectrum at the smallest wave-numbers $k$,
\begin{equation}
    b(M) = \sqrt{\frac{P_\text{\fastpm}(k)}{P_\text{Linear}(k)}} \,\Bigg|_{ k \,\rightarrow \,0} .
\end{equation}
We do this for two halo mass intervals: the one already used above $10^{12.3} \Mass  <M<10^{13.0} \Mass$ to show the power spectrum plots,  and in addition we choose $10^{13} \Mass  < M <10^{13.6} \Mass$. 
These simulated results are displayed with square  and  star  markers in Fig.~\ref{fig:halo-bias-k0p5-k1}, where the error bars denote the standard deviations of the 5 different realizations for each model. On the other hand, the filled circle markers in the plots show the theoretical results when averaged over the corresponding mass intervals, obtained using  Eq.~\ref{eq:effective_bias}. As it is clear from the plots, the match between simulations and theory is very good. As expected, the largest differences between the biases are located at smaller masses for the $k_T=1.0 \iMpc$ case (bottom panel) than for the  $k_T=0.5 \iMpc$ case (top panel). Further, the wider bumps, those with $\sigma_T=0.3$ (red dot-dashed lines), show larger deviations than the bumps with $\sigma_T=0.1$ (blue lines), which is expected because in the former case there is stronger gravitational interactions and consequently a greater clustering leading to a lower bias.

\subsection{Comparison to Knebe-Islam-Silk (2001)}\label{subsect:KIS}

In this subsection, we compare our results to the pioneering work of Knebe-Islam-Silk (KIS) from Ref.~\cite{Knebe:2001en}, as we have identified some overlaps with our research.

In the KIS study, the authors primarily focused on the cumulative mass function rather than the HMF. They reported some differences around the expected mass scales, but unfortunately, they did not show these differences explicitly. The crucial distinction between their work and ours lies in the fact that we directly present the halo mass function, clearly showcasing the appearance of the bump exactly where predicted by Sheth-Tormen. This is important because the cumulative mass function counts all masses below a certain range, and it is possible that an excess of halos in one range is compensated by a lack of halos in other ranges. This is particularly relevant when considering the Press-Schechter formalism (as in the KIS paper) or the Sheth-Tormen model (as in our work). Consequently, our research complements the results of KIS in the analysis of the HMF.

The only bump in KIS paper is the one they call ``Model 2'', the other models correspond to dips. This particular bump is located at $k=0.8 \, h \text{Mpc}^{-1}$ with a width of $\sigma_{mod}=0.1 \, h \text{Mpc}^{-1}$, which corresponds to $\sigma_T=0.16$ in our notation. Thus, the width and position of this bump are similar to those we study. However, KIS uses an amplitude of $A=3.0$, whereas our bumps are significantly smaller ($A=0.15$). Interestingly, they concluded that the bump is completely erased by evolution, which contradicts our findings indicating that the bump is somewhat preserved, albeit not entirely. We have observed the same effect in our previous paper \cite{Gomez-Navarro:2020fef} using different simulations and data analysis pipelines than those employed in this work. The results of KIS appear even more peculiar when considering that their bump amplitude is 20 times larger than ours. Furthermore, in Section V.B, we provide plausible physical explanations for why the bumps should be degraded by nonlinear evolution, and our results align with the theoretical expectations based on the Sheth-Tormen prescription for the HMF. (It should be noted that the simulations used in KIS utilized a significantly smaller number of particles, $128^3$, over boxes of size $256 \, h^{-1}\text{Mpc}$.)

Moreover, our discussion of halo bias differs from that of KIS, focusing on distinct aspects. In our work, we extensively discuss the large-scale, linear bias, providing theoretical reasons why it must differ from that in $\Lambda$CDM. This difference arises due to the significant changes in the variances $\sigma(M)$ in cosmologies featuring the bump. We even provide analytical estimates that remarkably agree with the simulated data. In contrast, KIS reports the existence of a different bias at very small scales, below $10\, h^{-1} \text{Mpc}$ (see Figure 10 of their paper), without providing explanations for its origin. In reality, we believe that some of these discrepancies can be attributed to our physically explained large-scale bias, which ultimately yields an offset in the overall amplitude of the power spectrum or correlation function. Therefore, the differences in the curves that are parallel in Figure 10 of KIS are likely mainly a consequence of the linear bias, as expected by peak-background-split theory. This conclusion is drawn from the results presented in our paper. Nevertheless, it is important to acknowledge that much more work needs to be done regarding nonlinear biasing.

In addition to the aforementioned differences, our study expands upon the following aspects:
1. We provide a more detailed analysis of the HMF by considering a Sheth-Tormen HMF with $p$ and $q$ free parameters. Although we ultimately conclude that these parameters are consistent with the standard values, which emphasize the universality of Sheth-Tormen  beyond LCDM). 2. We offer a detailed analysis of the halo abundance, even providing histograms for the different models. 3.  Based on the halo model, we provide theoretical explanations for the degradation of the bump in the matter power spectrum. 4. We confirm the existence of a second bump, which has not previously been reported in the literature, and provide heuristic reasoning supporting its presence.

\section{Conclusions}\label{section:conclusions}

Bump cosmologies are inspired by models beyond $\Lambda$CDM that have dark sector energy densities that suffer phase transitions, leaving distinctive features in abundance and the clustering data due to a Rapid Diluted Energy Density (RDED). In such scenarios, the power spectrum is enhanced at scales where otherwise the power would be smooth, originating a primordial bump at nonlinear scales relative to a model with no phase transition. 
%
%
This can be understood since the presence of the energy density that had a transition from being relativistic to behave as a fluid with an equation of state parameter $w=1$.  
As a result, this additional energy density $\rho_{ex}$ dissipates faster than radiation for $a>a_T$  having an impact on the expansion rate parameter $H$ and noticeable in the ratio  $H_\text{$\Lambda$CDMex}(a)/H_\text{$\Lambda$CDM}(a)$ impacting also the growth of structure and generating a bump in the ratio matter power spectra $P_\text{$\Lambda$CDMex}(a)/P_\text{$\Lambda$CDM}(a)$.


In this work, we have studied halo abundance and clustering in bump cosmologies. Instead of considering any specific BDE model, we have used a parametric family of bumps, allowing us to explore a wider range of theoretical models. We have run \fastpm N-body simulations \cite{Feng:2016yqz}, which are complemented by nonlinear halo model approximations from the \textsc{HMcode} model \cite{Mead2015}. We noticed that the nonlinear effects shift the peaks and originates a second bump at smaller scales because the primordial, original linear bumps serve as regions where average densities are higher than the background, the gravitational collapse becomes more rapid and efficient, and peaks can cross the critical density threshold for collapse more often. We have focused on the abundance and clustering statistics and mainly on the responses as given by the ratio of summary statistics in a bump cosmology to a $\Lambda$CDM cosmology with no bump. This analysis is useful since $\Lambda$CDM is well known, and then good models for the response translate into good models for the statistics themselves. We have studied the nonlinearities in the matter and halo power spectrum and how these fingerprints are translated to large-scale halo bias. We also have studied how the number of halos are affected by the phenomenology of the bump cosmology.

We have analysed and compared to $\Lambda$CDM, four bump parametrized cosmologies. These are the combinations of two locations $k_T=0.5$ and $1 \iMpc$ and two widths $\sigma_T=0.3$ and $0.1$. The location of the bump corresponds to the mode reentering the horizon at the phase transition redshift, while the width to the duration of the phase transition. 

We have confirmed that the power spectrum in the RDED cosmologies is affected by two important nonlinear effects. First, the bumps are erased because the large scale random bulk matter motions tend to populate underdense regions, while moving out of regions with less matter than the average. This effect has a similar origin than the BAO damping and is more evident for bumps located at higher $k$ modes because these translate in oscillations in configuration space with smaller distances between crests and troughs. The second effect is the appearance of a second bump due to the fact that the primordial one serves as a location with high density and where even smaller structures are more prone to form.

We have seen that the presence of localized bumps in the matter power spectrum have consequences on the halo statistics at all scales. The halo power spectrum suffers an offset with respect to the $\Lambda$CDM one because the large scale bias is sensitive to the variance of density fluctuations $\sigma(M)$, which is affected by the bump. The differences with the $\Lambda$CDM can be considerable even for small bumps located well inside the nonlinear region. 

We have computed the Halo Mass Function (HMF) for our bump cosmologies using the Sheth-Tormen recipe. However, while the Sheth-Tormen HMF has fixed parameters, we fit them to the simulations anticipating the possibility of more optimal parameters for bump cosmologies. However, we found their values to be very close to the standard $p=0.3$ and $q=0.707$. Despite this, the HMFs do differ largely for the bump and $\Lambda$CDM cosmologies, but this is because of different matter power spectra and then different variances $\sigma(M)$. We compared our analytical results to the outcomes of the N-body simulations finding excellent agreement. Further, we were capable of correctly predict the large scale halo bias by applying the Peak-Background-Split formalism to our Sheth-Tormen HMF description, matching the simulated power spectrum data at low $k$ for two ranges of masses: $10^{12.3} \Mass <M<10^{13.0} \Mass$ and $10^{13} \Mass < M <10^{13.6} \Mass$. 

In general, for the bump cosmologies and within the adequate mass ranges, that is, within the ranges where the formation of halos of certain masses are more enhanced, we found smaller biases than in $\Lambda$CDM. This is because in those regions the presence of the primordial bump yields more clustering and then stronger gravity effects, which translate in a more efficient relaxation of the bias, which tends toward unity more rapidly than in a cosmology without the bump. 

In summary, this distinctive fingerprint, named as a bump, has been studied modelling the statistics of biased tracer of the density, such as halos, which have the potential to be detectable by current and future galaxy surveys \cite{Conroy_2005, NorbergFrenkCole2007, Hudson2014, MandelbaumWangYingWhite2016} allowing to put tight constraints on cosmological constraints. Future interesting routes to continue the analysis of this work include the investigation of the halo-galaxy connection and the clustering in mock galaxy catalogs that can shed light on the expected observables for galaxies surveys such as the Dark Energy Spectroscopic Instrument (DESI) \cite{2016arXiv161100036D}. Also, since weak lensing depends critically on the real-space matter power spectrum, whose nonlinear effects are easy to understood and well modeled by the halo-based \textsc{HMcode}, we forecast that the consequences of bump cosmologies on statistics measured by photometric surveys (e.g. the $3\times 2$ point correlation functions) are relatively straightforward to obtain, with a view on the upcoming Rubin Observatory Legacy of Space and Time (LSST) Survey \cite{2012arXiv1211.0310L}.

\section*{Acknowdledgments}
DGN thanks support from a CONACyT PhD fellowship. AM and DGN acknowledges  support from PAPIIT- DGPA (UNAM) IN101415. A.~A. is supported by Ciencia de Frontera grant No.~319359, and also acknowledges partial support to grants Ciencia de Frontera 102958 and CONACyT 283151.

\bigskip

\appendix

\begin{section}{Halos per mass interval}\label{app:Nhalos}


In Fig.~\ref{fig:histogram-halo-mass}, we show histograms for the counts of halos as a function of their mass within the interval $M=10^{12.3} \Mass$ to $M=10^{15} \Mass$. Complementary to this figure, in Table \ref{tab:mean-number-halo-catalog} we show the mean number of halos of our catalog suite. Both in the Figure and in the Table, the errors show the standard deviations over the 5 simulations for each model. At higher redshift, the mean number of halos in bump cosmologies is larger than in $\Lambda$CDM. Meanwhile, at present time the mean number of halos of $\Lambda$CDM is larger than bump cosmologies for the cases $k_T=0.5 \iMpc$, but not for $k_T=1.0 \iMpc$ cases.

\begin{figure*}
\includegraphics[height = 0.27\textheight]{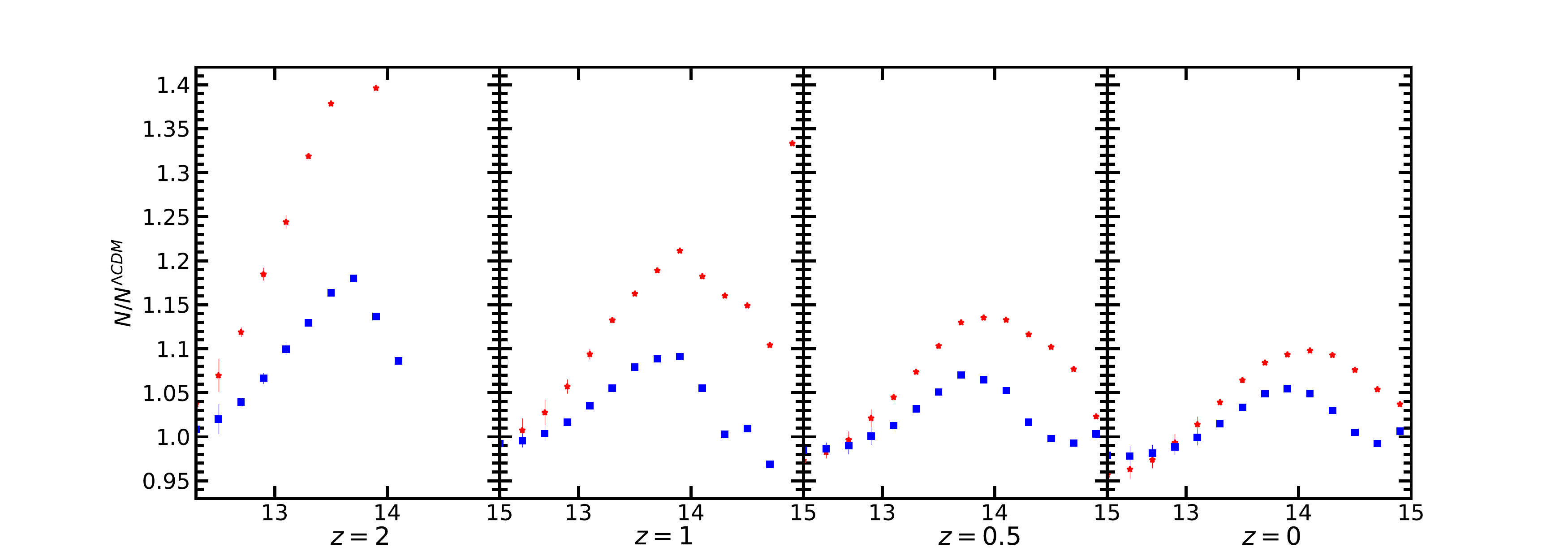}
\includegraphics[height = 0.27\textheight]{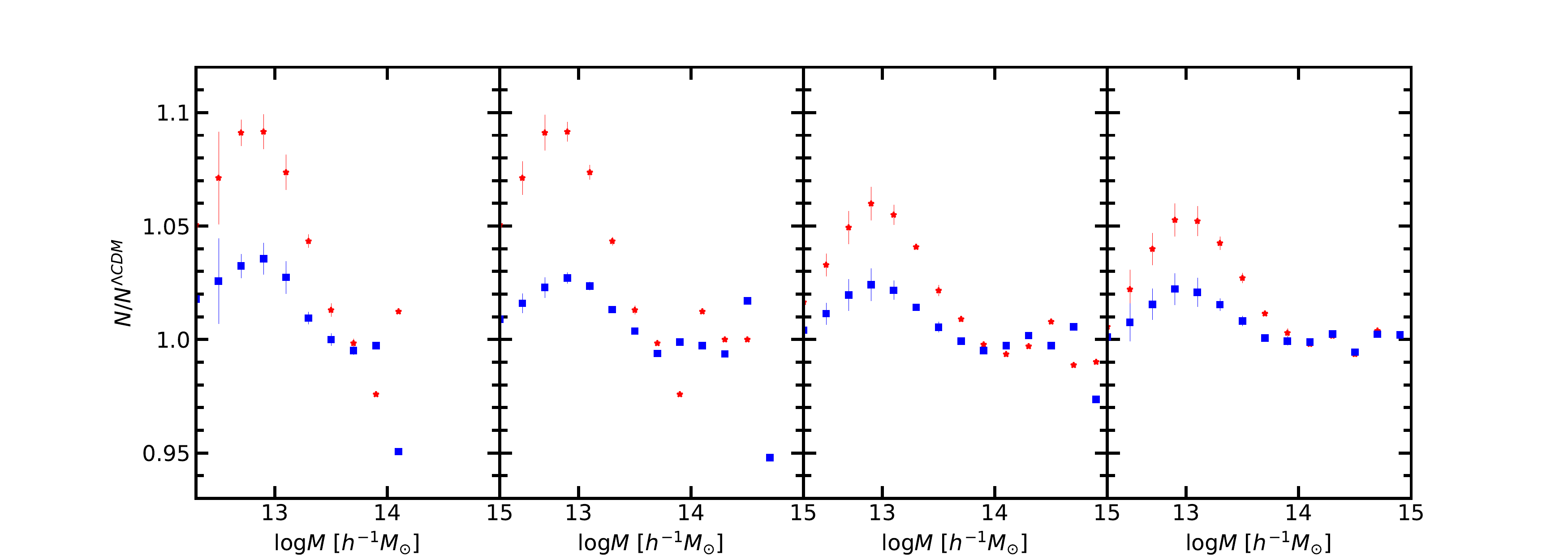}
\caption{Histogram of halos for the bump cosmologies at $k_T=0.5 \iMpc$ (top panel) and $k_T=1 \iMpc$ (bottom panel). Stars are for the measurement from N-body simulations with $\sigma_T=0.3$; and squares for $\sigma_T=0.1$. Error bars that are not visible are within the mark size.}
\label{fig:histogram-halo-mass}
\end{figure*}

 \begin{table*}
	\centering
	\begin{tabular}{l c c c c } 
		\hline
		\\ [-3pt]
		Name $\qquad$ & $\qquad$  $\overline{N}(z=2.0)$ $\qquad$ & $\qquad$ $\overline{N}(z=1.0)$ $\qquad$ & $\qquad$ $\overline{N}(z=0.5)$ $\qquad$ & $\qquad$ $\overline{N}(z=0.0)$ $\qquad$ \\ [2pt]
		\hline
		\\ [-3pt]
		\textsc{medbump-k1} & $870 050 \pm 1239$ & $1 703 776 \pm 2872$ & $2 116 618 \pm 3923$ & $2 382 548 \pm 4320 $ \\[4pt]
		\textsc{thinbump-k1} & $835 498 \pm 1538$ & $1 658 230 \pm 3460$ & $2 072 237 \pm 4096$ & $2 344 930 \pm 4590$ \\[4pt]
		\textsc{medbump-k0p5} & $882 950 \pm 1468$ & $1 673 741 \pm 2985$ & $2 048 335 \pm 4240$ & $2 280 528 \pm 3735$ \\[4pt]
		\textsc{thinbump-k0p5} & $837 524 \pm 1574$ & $1 640 229 \pm 3557$ & $2 037 594 \pm 4161$ & $2 294 898 \pm 4134$ \\[4pt]
		\textsc{$\Lambda$CDM}& $816 183 \pm 1639$ & $1 632 756 \pm 3499$ & $2 047 768 \pm 4369$ & $2 323 036 \pm 4598$\\[4pt]
		\hline
	\end{tabular}
	\caption{Mean number of halos of our halo catalog suite. Column I corresponds to the models' names; column II to mean number $\overline{N}$ at $z=2$; column III to $z=1$; column IV to $z=0.5$; and column V to $z=0$. 
 }
	\label{tab:mean-number-halo-catalog}	
\end{table*}

\end{section}

\bibliography{biblio}

\end{document}